\documentclass[aps,pra,twocolumn,groupedaddress,floatfix,showpacs,longbibliography]{revtex4-1}
\usepackage{graphicx}
\usepackage{amsmath}
\usepackage{color}
\def\ket#1{|#1\rangle}
\def\bra#1{\langle#1|}

\begin{document}
\title{Verifying a quasi-classical spin model of perturbed quantum rewinding in a Fermi gas}

\author{J. Huang, Camen A. Royse, I. Arakelyan, and J. E. Thomas}

\affiliation{$^{1}$Department of Physics, North Carolina State University, Raleigh, NC 27695, USA}

\date{\today}

\begin{abstract}
We systematically test a quasi-classical spin model of a large spin-lattice in energy space, with a tunable, reversible Hamiltonian and effective long-range interactions. The system is simulated by a weakly interacting Fermi gas undergoing perturbed quantum rewinding using radio-frequency(RF) pulses. The model reported here is found to be in a quantitative agreement with measurements of the ensemble-averaged energy-resolved spin density. This work elucidates the effects of RF detunings on the system and measurements, pointing the way to new correlation measurement methods.

\end{abstract}

\maketitle

Measurement of coherence, entanglement, and correlations in time-reversible many-body spin lattices is of great interest, broadly impacting our understanding of quantum measurement and information processing~\cite{OTOCNatComm, QMBNatPhy, QEntangleScience}. A nearly ideal platform for simulating large spin-lattices is a weakly interacting Fermi gas, containing $N\simeq 10^5$ atoms with a tunable, reversible Hamiltonian. The trapped cloud behaves as a spin lattice in energy-space with effective long-range interactions~\cite{DuSpinSeg1,DuSpinSeg2,LewensteinDynLongRange,SaeedPRASpinECorrel,Piechon,MuellerWeaklyInt,LaloeSpinReph,
ThywissenDynamicalPhases,KollerReySpinDep,ReyNatPhys2017,OutofEq} and enables new tests of classical versus quantum spin evolution~\cite{Schubert,Kulkarni,Lakshmanan, CLvsQPhysWorld}.

Spin waves are observed in weakly interacting, nearly collisionless Fermi gases, which have been explained by several models~\cite{LewensteinDynLongRange,DuSpinSeg2,Piechon,MuellerWeaklyInt,LaloeSpinReph,KollerReySpinDep,SaeedPRASpinECorrel}. Previously, a 1D quasi-classical spin evolution model that uses the exact energy-dependent couplings was found to yield agreement with spin-density profiles measured for the evolution of an initially $x$-polarized spin sample~\cite{SaeedPRASpinECorrel}. However, it appeared that this model failed to explain perturbed quantum rewinding experiments, where an RF pulse rotates the entire spin system by an angle $\phi_x$ about the $x$ axis as a perturbation in between forward and backward evolutions. In a quantum picture, the $\phi_x$ rotation changes the relative phases of the superposed total angular momentum states that describe the system, i.e., $|S,M_x\rangle\rightarrow e^{-iM_x\phi_x}|S,M_x\rangle$ for each state, leading to $\phi_x$-dependent coherence amplitude between states differing in $M_x$~\cite{ReyNatPhys2017,ReyPRL2018,SaeedInformScramb}. To fit the data, a scattering length of $\approx2.5$ times the measured value was needed in the previous work~\cite{SaeedInformScramb}, questioning the adequacy of the quasi-classical model.

In this work, we report precise, systematic tests of a modified quasi-classical spin model using single-shot measurements of the spin density profiles from perturbed quantum rewinding experiments. Such experiments are ideal for testing the model, since unperturbed rewinding experiments can be implemented in advance to prove that the system is reversed properly without model-dependent fits. We show the advantages of single-shot data analysis for studies of ensemble-averaged energy-resolved spin density, and quantitatively demonstrate the important roles of different RF detunings during the forward and backward evolution periods. By using two detunings as separate fit parameters, the data is explained by the model using the measured scattering length. The new approach reported here validates the modified quasi-classical treatment of this quantum spin system and suggests detuning-independent measurement methods for future correlation studies, avoiding probabilistic methods in data selection~\cite{ThywissenDynamicalPhases}.

Our experiments~\cite{Supplement}, employ degenerate clouds of $^6$Li containing a total of $N=6.5\times 10^4$ atoms initially in a single spin state. The cloud is confined in a harmonic, cigar-shaped optical trap, with oscillation frequencies $\omega_x/2\pi=24.4$ Hz in the axial direction and $\omega_r/2\pi=650$ Hz in the radial direction. The corresponding Fermi temperature $T_F=0.73\,\mu$K and $T/T_F=0.31$. RF pulses prepare coherent superpositions of the two lowest hyperfine-Zeeman states, which are denoted by $\ket{1}\equiv\ket{\!\uparrow_z}$ and $\ket{2}\equiv\ket{\!\downarrow_{z}}$. The experiments are done in the weakly interacting regime, where the energy-state changing collision rate is negligible over the time scale of the measurements~\cite{SaeedPRASpinECorrel}.

As the single particle energies are {\it fixed} and the energy distribution is time independent~\cite{SaeedPRASpinECorrel}, we approximate the cigar-shaped weakly interacting Fermi gas as a one-dimensional (1D) spin ``lattice" in energy space~\cite{SaeedPRASpinECorrel}, with a Hamiltonian
\begin{equation}
\frac{H(a)}{\hbar}=a\!\sum_{i,j\neq i}\!g_{ij}\,{\vec s}_i\cdot{\vec s}_j+\sum_{i}\Omega'E_i\,s_{zi}+\Delta(t)S_{z}.
\label{eq:1.1}
\end{equation}
\begin{figure*}[htb]
\begin{center}\
\hspace*{-0.25in}\includegraphics[width=5.75in]{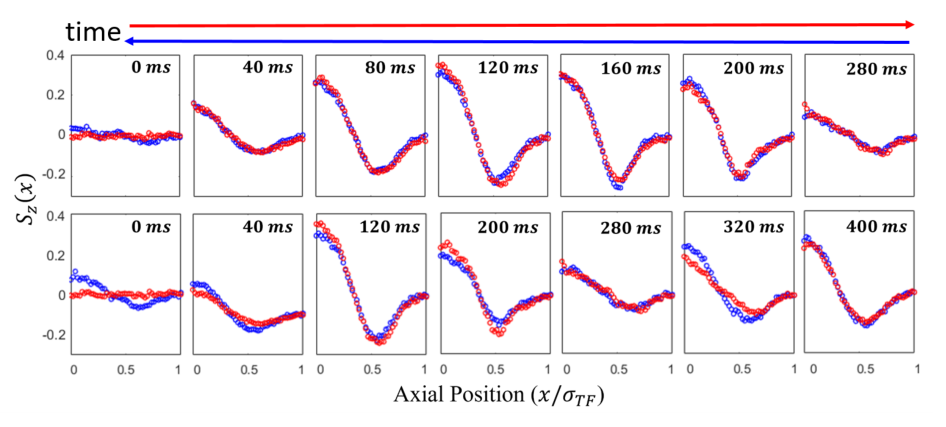}
\end{center}
\caption{Observing quantum rewinding by comparing the spin densities for forward evolution (red circles) and backward evolution (blue circles). The scattering length $a=8.0\,a_0$ and $\sigma_{TF}\approx 335\,\mu$m. The Hamiltonian is reversed at $\tau_f=280$ ms (top) and $\tau_f=400$ ms (bottom). For forward evolution data, the time $t_k$ shown on the top right corner of each tile is the total evolution time $t_{\rm fk}$: For the backward evolution data, the total evolution time $t_{\rm bk} = \tau_f + \tau_b$, where $\tau_f$ is the time at which reversal occurs, and $\tau_b = \tau_f - t_k$ is the backward evolution time, see Fig.~\ref{fig:OTOC}.\label{fig:rewind} To avoid confusion with the $x$-axis of RF or Bloch frame, here, we clarify that $x$ in all spatial profiles means axial direction along the longitudinal axis of the cigar-shaped cloud.}
\end{figure*}
We associate a ``site" $i$ with the energy $E_i\!=\!(n_i\!+\!1/2)\,h\nu_x$ of the i$^{\rm th}$ harmonic oscillator state along the cigar axis $x$. For each $E_i$, we define a dimensionless collective spin vector $\vec{s}\,(E_i)\equiv\vec{s}_i$.

The first term in Eq.~\ref{eq:1.1} is the site-to-site interaction, proportional to the s-wave scattering length $a$ and to the overlap of the harmonic oscillator probability densities for colliding atoms. In a WKB approximation, $g_{ij}\propto 1/\sqrt{|E_i-E_j|}$, which is an effective long-range interaction in the energy-space lattice~\cite{SaeedPRASpinECorrel}. For a zero temperature Fermi gas, the average interaction energy (in rad/s) is $a\bar{g}=6.8\,n_0\hbar a/m$, where $n_0$ is the peak density. For our experimental parameters, with $a=5.2\,a_0$, $a\bar{g}/2\pi\simeq 2.0$ Hz.

The second term in Eq.~\ref{eq:1.1} is an effective site-dependent Zeeman energy, arising from the quadratic spatial variation of the bias magnetic field along $x$, which produces a spin-dependent harmonic potential. As $\omega_r/\omega_x=26.6$, the corresponding effect on the radial motion is negligible, enabling a 1D approximation, where all atoms at site $i$ have the same Zeeman energy. In Eq.~\ref{eq:1.1}, $\Omega'=-\delta\omega_x/(\hbar\omega_x)$, with $\delta\omega_x/2\pi=14.9$ mHz for our trap~\cite{SaeedPRASpinECorrel}. For the mean energy $\bar{E}_x\simeq k_B T_F/4$, $\Omega'\,\bar{E}_x/2\pi\simeq 2.0$ Hz.

The last term in Eq.~\ref{eq:1.1} arises from the time-dependent global detuning $\Delta(t)$, which plays a central role in the analysis of the rewinding data. Here, $S_z=\sum_i s_{zi}$. For a typical evolution time $~200$ ms, $\Delta(t)\simeq0.4$ Hz. Fluctuations in the bias magnetic field and magnetic tuning of the scattering length cause $\Delta(t)$ to change at 5 kHz/G for $|1\rangle$-$|2\rangle$ superposition states.

\begin{figure}[htb]
\begin{center}\
\hspace*{-0.0in}\includegraphics[width=3.0in]{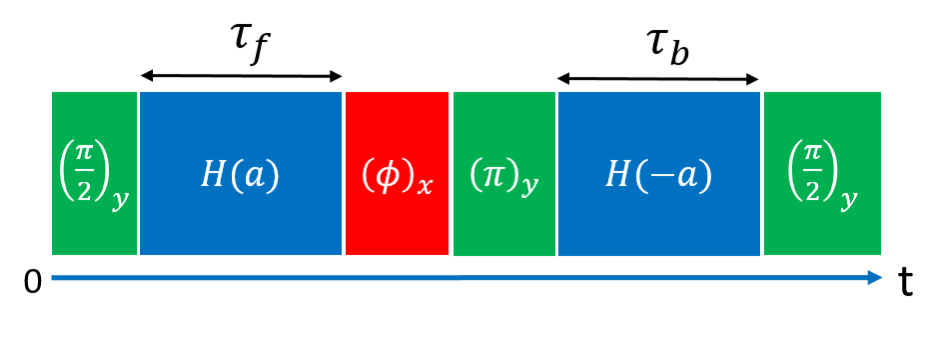}
\end{center}
\caption{Characterizing perturbed quantum rewinding. The atom cloud is initially prepared in a pure $z$-polarized spin state. The system runs forward for a time $\tau_f$ and backward for $\tau_b$, after which the spatial profiles of the $|\uparrow_z\rangle$ and $|\downarrow_z\rangle$ states are measured. Green and red blocks represent RF pulses and rotation angles about the $y$- and $x$-axes respectively, whose durations are $<<\tau_f,\,\tau_b$.
\label{fig:OTOC}}
\end{figure}

To implement perturbed quantum rewinding, we employ the pulse sequence shown in Fig.~\ref{fig:OTOC}~\cite{SaeedInformScramb}. The system is initially prepared in a pure z-polarized spin state, $\ket{\psi_{0z}}$. The first $(\pi/2)_y$ pulse (0.5 ms), defined to be about the $y$-axis, creates an $x$ polarized state, $\ket{\psi_{0x}}$. Here, the $y$- and $x$-axes are defined in the rotating frame of the RF pulses (RF-frame). Then, the system is allowed to evolve forward for a time $\tau_f$. A voltage-controlled change of the RF phase by $\pi/2$ permits rotation about the $x$-axis by an angle $\phi_x$. Applying a $(\pi)_y$ pulse (1 ms) and magnetically tuning the scattering length from $a\rightarrow -a$ (10 ms) inverts the sign of Hamiltonian shown in Eq.~\ref{eq:1.1}, causing the system to evolve backward for a time $\tau_b$ \cite{Supplement}. As described below, we perform experiments both with and without the final $(\pi/2)_y$ pulse, after which the spatial profiles of the $|\uparrow_z\rangle$ and $|\downarrow_z\rangle$ states are measured by two resonant absorption imaging pulses, separated by $10\,\mu$s, to obtain the single-shot spin density $S_z(x)=[n_{\uparrow_z}(x)-n_{\downarrow_z}(x)]/2$. For each shot, $S_z(x)$ is normalized to the total central density $n(0)=n_{\uparrow_z}(0)+n_{\downarrow_z}(0)$ to minimize errors arising from shot-to-shot variation in the atom number and cloud width. All spatial profiles are folded about $x=0$ and displayed for $0\leq x\leq\sigma_{TF}$.

The reversibility of the system is tested (result shown in Fig.~\ref{fig:rewind}) using the pulse sequence of Fig.~\ref{fig:OTOC} with
$\phi_x=0$ and {\it without} the final $(\pi/2)_y$ pulse. This sequence measures the component of the collective spin vector ${\vec s}_i$ that was along the $z$-axis just prior to imaging. The longitudinal ($z$) component is insensitive to the detuning $\Delta(t)$ that causes a rotation of ${\vec s}_i$ about the $z$-axis relative to the RF-frame, enabling a robust test. In the data analysis, since $S_z=0$ for $\phi_x =0$, global spin balance is enforced to minimize the error from small shot-to-shot changes in the detuning of the RF pulses, which arises from magnetic field fluctuation.

In these experiments, it is essential to carefully calibrate the bias magnetic field $B_0$ at which the s-wave scattering length vanishes. This is best done by quantifying the reversal results using different magnetic fields, which is independent of fitting models and less sensitive to the initial conditions in contrast to the method adopted in Ref.~\cite{SaeedPRASpinECorrel}. $B_0$ is found by minimizing the sum of the mean square differences between the forward and backward spin density profiles at corresponding times~\cite{Supplement}. Unperturbed rewinding experiments done at scattering lengths of $\pm 5.2\,a_0$ and $\pm 8.0\,a_0$, suggest that $B_0=527.150(5)$ G, which is lower by $30$ mG compared to the $B_0$ of Ref.~\cite{SaeedPRASpinECorrel}.

Fig.~\ref{fig:rewind} shows rewinding data (6-shot average) at corresponding forward(red) and backward(blue) evolution times for $a=8.0\,a_0$ and $-8.0\,a_0$ respectively. With the calibrated $B_0$, the corresponding forward and backward spin density profiles demonstrate good agreement for reversal at $280$ ms (top row), while reversal at 400 ms (bottom row) leads to greater differences between corresponding forward and backward data profiles.
\begin{figure}[htb]
\includegraphics[width=3.25in]{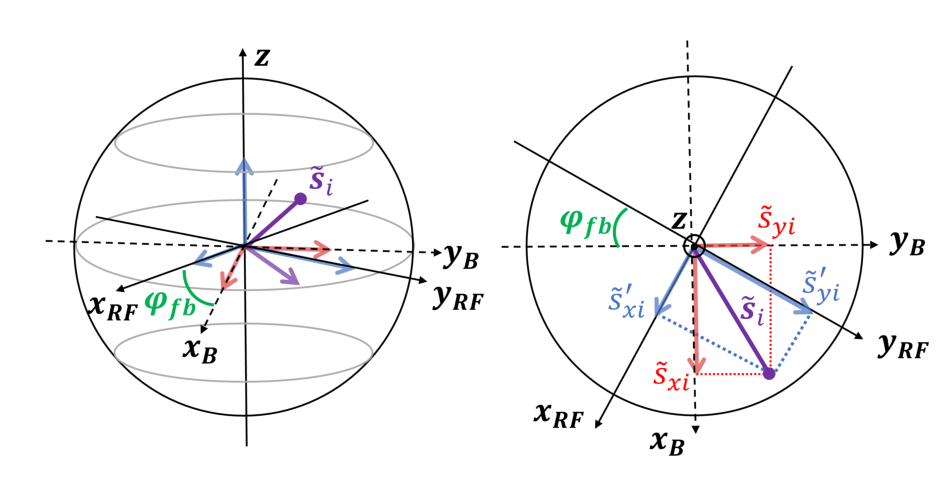}
\caption{Relation between the spin vector components in the radiofrequency(RF) and Bloch(B) frames for nonzero $\varphi_{fb}$.
\label{fig:BlochSphere}}
\end{figure}

\begin{figure*}[htb]
\begin{center}\
\hspace*{-0.25in}\includegraphics[width=5.75in]{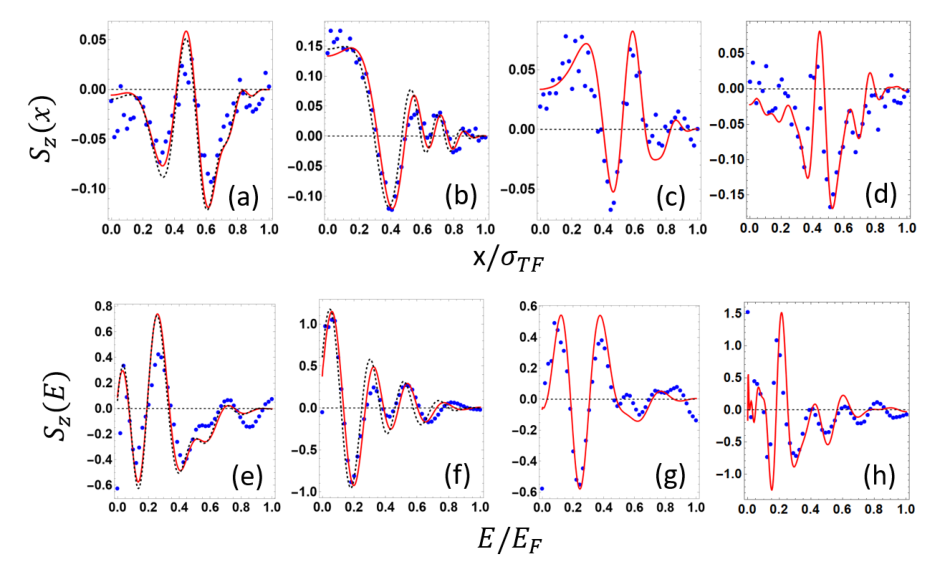}
\end{center}
\caption{Single-shot spin-density profiles in space and in energy space for perturbed quantum rewinding. For this set of data, $\sigma_{TF}\approx340\, \mu$m, (a,e) measured scattering length $a=5.2\,a_0$, $\phi_x=\pi/2$, $\tau =200$ ms; (b,f) $a=5.2\,a_0$, $\phi_x=\pi$, $\tau =200$ ms; (c,g) $a=8.0\,a_0$, $\phi_x=3\pi/2$, $\tau =200$ ms; (d,h); $a=5.2\,a_0$, $\phi_x=\pi/2$, $\tau =400$ ms. Blue dots are single-shot data. Red curves: Quasi-classical collective spin vector model using the measured scattering length, with forward and backward evolution phases $\varphi_f$ and $\varphi_b$ as fit parameters. Black-dashed curves show the fits with $\varphi_b=\varphi_f$ as one free parameter and scattering length $a_{fit}$ as the other, which requires $a_{fit}=9.0\,a_0$.
\label{fig:OTOCPhix}}
\end{figure*}

Having established that the system is reversible for scattering lengths up to $\pm 8.0\,a_0$ and $\tau_f=\tau_b\leq 280$ ms, data are mainly obtained with $\tau\equiv\tau_f=\tau_b =200$ ms at $\pm 5.2\,a_0$ using the full pulse sequence of Fig.~\ref{fig:OTOC}. This provides stringent tests of quasi-classical collective spin vector models. Here, the final $(\pi/2)_y$ pulse is included to measure the transverse spin components that were along the $x$-axis in the RF frame in Fig.~\ref{fig:BlochSphere} just prior to imaging. For $\phi_x=0$ and a detuning $\Delta(t)$ that is constant over the total sequence, the system is expected to rewind to the initial state, where the density profiles for both spins are Thomas-Fermi. For $\phi_x\neq 0$, however, the rewinding is perturbed, producing complex spin density profiles after the full sequence. Fig.~\ref{fig:OTOCPhix} shows single-shot spin density profiles for $\phi_x=\pi/2,\pi,3\pi/2$. We obtain the corresponding energy-space profiles, $ s_{zi}\equiv s_z(E)$, by inverse Abel-transformation \cite{NewAbelInversion} of the spatial profiles, which is valid in a WKB approximation when energy space coherence is negligible and a quasi-continuum approximation is valid, as in our experiments~\cite{SaeedPRASpinECorrel}.

To understand the perturbed rewinding data of Fig.~\ref{fig:OTOCPhix}, we include a time-dependent global detuning, $\Delta(t)$, in the Hamiltonian of Eq.~\ref{eq:1.1}. The detuning determines the relative angle between the RF-frame and the Bloch frame $\varphi_{fb}$ in Fig.~\ref{fig:BlochSphere}. Here, the RF frame is defined by $x_{RF}$ and $y_{RF}$ axes that rotate about the $z$-axis at the instantaneous RF frequency, $\omega_{RF}(t)$, tracking the {\it total} phase of the RF field. We define the rotation axes for all of the RF pulses in Fig.~\ref{fig:OTOC} to be in the RF frame, i.e., $x\equiv x_{RF}$ and $y\equiv y_{RF}$. The Bloch frame is defined by $x_B$ and $y_B$ axes that rotate at the instantaneous hyperfine resonance frequency $\omega_{HF}(t)$ for an atom of axial energy $E=0$.

The detuning, $\Delta(t)=\omega_{HF}(t)-\omega_{RF}(t)$, causes the components of spin vectors in the Bloch frame to rotate relative to the RF-frame by generally different angles $\varphi_f=\int_{\tau_f}\!\! dt\,\Delta(t)$ and $\varphi_b=\int_{\tau_b} \!\!dt\,\Delta(t)$, during the forward and backward evolution times, respectively, even for $\tau_b=\tau_f$. For measurements of spin components in the RF frame, the final state of the cloud can be written as $\ket{\psi_f}=e^{-i\pi/2\, S_y}\ket{\psi_{f1}}$, where
$\ket{\psi_{f1}}$ is the state just prior to the final $(\pi/2)_y$ pulse. Taking $\tau_f=\tau_b=\tau$, we find~\cite{Supplement}
\begin{equation}
\ket{\psi_{f1}}=e^{-i\pi S_y}e^{i(\varphi_b-\varphi_f)S_z}\,W_\phi(\varphi_f,\tau)\ket{\psi_{0x}},
\label{eq:statef1}
\end{equation}
where $\ket{\psi_{0x}}=e^{-i\pi/2\, S_y}\ket{\psi_{0z}}$ is the fully $x$-polarized state and
\begin{equation}
W_\phi(\varphi_f,\tau)=e^{\frac{i}{\hbar}H_0(a)\tau}e^{-i\,\phi_x S_x(\varphi_f)}e^{-\frac{i}{\hbar}H_0(a)\tau}.
\label{eq:2.9}
\end{equation}
Here, $S_x(\varphi_f)=S_x\cos\varphi_f-S_y\sin\varphi_f$ with $S_x$ and $S_y$ the $x$- and $y$-components of the {\it total} spin vector in the RF frame. $H_0(a)$ is defined by Eq.~\ref{eq:1.1} for $\Delta(t)=0$.

For each shot, the operator $s_{zi}$ is measured for an ensemble of atoms in a selected energy group $E_i\in [E,E+\Delta E]$. The energy resolution $\Delta E$ of the inverse Abel-transform method is small enough that all of the atoms in the energy group evolve identically over the time scale of the pulse sequence. A single-shot measurement of the spin density profile then yields the ensemble-averages, $\langle\psi_f|s_{zi}|\psi_f\rangle \equiv\bra{\psi_{0x}}\tilde{s}'_{xi}\ket{\psi_{0x}}$. Here,
\begin{equation}
\tilde{s}'_{xi}= \cos\varphi_{fb}\,\tilde{s}_{xi}-\sin\varphi_{fb}\,\tilde{s}_{yi},
\label{eq:3.9main}
\end{equation}
with $\tilde{s}'_{xi}$ being the $x$-component of the spin vector operator relative to the RF frame just before the final $(\pi/2)_y$ pulse as shown in Fig. \ref{fig:BlochSphere}. $\tilde{s}'_{xi}$ is given in terms of the components in the Bloch frame, $\tilde{s}_{xi}\equiv W^\dagger_\phi(\varphi_f,\tau)s_{xi}W_\phi(\varphi_f,\tau)$ and similarly for $\tilde{s}_{yi}$. For each measurement, we see that the difference between the backward and forward the phase shifts, $\varphi_f-\varphi_b\equiv\varphi_{fb}$, determines the relative contribution of the $\tilde{s}_{xi}$ and $\tilde{s}_{yi}$ spin components in the Bloch frame to the measured projection in the RF frame, $\tilde{s}'_{xi}$. In addition, Eq.~\ref{eq:2.9} shows that the forward phase shift $\varphi_f$ determines the effective rotation axis for the $\phi_x$ pulse.

To predict the measured $\langle\psi_f|s_{zi}|\psi_f\rangle$, we employ a mean-field approximation to obtain a quasi-classical model~\cite{SaeedPRASpinECorrel}, where the Heisenberg equations are solved numerically by treating the collective spin vectors as classical variables, which ignores quantum correlations between the spin vectors for different energy groups. The Heisenberg equations of motion for the collective spin vectors take a simple form in energy space, $\dot{\vec{s}}_i(t)=\vec{\omega}_i(t)\times{\vec{s}}_i(t)$, with
\begin{equation}
\vec{\omega}_i(t)=a\!\sum_{j\neq i}g_{ij}\,{\vec s}_j (t)+\Omega'E_i\,\hat{e}_z+\Delta(t)\hat{e}_z.
\label{eq:omega}
\end{equation}
For a given choice of the forward and backward detunings, i.e., the phases $\varphi_f$ and $\varphi_b$, $s_{zi}$ is determined by numerical integration. An Abel transform of $s_{zi}\equiv s_z(E)$ then yields the corresponding spin density $s_z(x)$~\cite{SaeedPRASpinECorrel}.

Experimentally, 60 shots are taken for each set of parameters. Examples of single-shot data are shown in Fig.~\ref{fig:OTOCPhix} and in the supplement \cite{Supplement}. Due to the complexity of the spatial profiles for $\phi_x\neq0$, single-shot data analysis is essential for this experiment. Small variation ($\leq5\%$) in cloud parameters results in shifted spatial profiles, even for fixed $\varphi_f$ and $\varphi_b$, so averaging over shots with slightly different initial conditions can wash out the fine structure. Fig.~\ref{fig:OTOCPhix} compares two quasi-classical models with the single-shot data (blue dots). The first model, adopted from Ref.~\cite{SaeedInformScramb}, assumes $\varphi_f\equiv\varphi_b$ mod $2\pi$, and the fits (black-dashed curves) to the data in Fig.~\ref{fig:OTOCPhix} (a,e) and (b,f) for $\tau=200$ ms and $a=5.2\,a_0$ require a {\it fitted} scattering length of $a_{fit}=9.0\,a_0$ in disagreement with the measured value. These results confirm the large discrepancy between the data and the quasi-classical model ignoring $\varphi_{fb}$ that was observed in our previous study of information scrambling~\cite{SaeedInformScramb}. For the second, modified model, the forward and backward evolution phases $\varphi_f$ and $\varphi_b$ are treated as two free parameters. In this case, the model (red curves) is in good agreement with data taken for $\phi_x=\pi/2$, $\pi$ and $3\pi/2$ with $\tau=200$ ms at both $5.2\,a_0$ and $8.0\,a_0$ and for $\tau=400$ ms at $5.2\,a_0$. Additional data with $\phi_x$ in steps of $\phi_x=\pi/4$ were obtained to test the model further and demonstrate equally good agreement~\cite{Supplement}. Section IV B of the supplement explains the sources of minor defects observed in the data series for $8.0\,a_0,~\tau=200$ ms and $5.2\,a_0,~\tau=400$ ms.

It is observed that, for the small scattering length, $a=5.2\,a_0$, and short forward evolution time $\tau=200$ ms, the data can be fitted using $\varphi_f$ and $\varphi_b$ as two free parameters (red curves) or by using $\varphi_f=\varphi_b$ as one parameter and the scattering length $a$ as another free parameter (black-dashed curves). However, for the long forward evolution time of 400 ms or for the large scattering length of $8.0\,a_0$, the data cannot be fitted for any scattering length with the assumption of $\varphi_b=\varphi_f$. In contrast, the modified model reported in this work, which includes forward and backward evolution phases as separate parameters, fits the data very well using the measured scattering length.

The modified model explicitly shows the difficulty of multi-shot averaged measurements of transverse spin components, such as $s_x$, where the averages of $\cos\varphi_{fb}$ and $\sin\varphi_{fb}$ in Eq.~\ref{eq:3.9main} tend to vanish. Previously, the imperfect phase control problem was partially circumvented by using a maximum likelihood estimation~\cite{ThywissenDynamicalPhases}. However, Eq.~\ref{eq:3.9main}, which is valid for both quasi-classical and full quantum treatments, suggests that multi-shot averaged measurements of energy-space spin operator products, such as $\langle\psi_f|s_{zi}s_{zj}|\psi_f\rangle =\bra{\psi_{0x}}\tilde{s}'_{xi}\tilde{s}'_{xj}\ket{\psi_{0x}}$, are important, since the random-phase averages of $\cos^2\varphi_{fb}$ and $\sin^2\varphi_{fb}$ are nonzero. This method enables improved out-of-time-order correlation measurements in quantum gases, where the $W$ operator is unchanged and the operator $V=s_{xi}$ is replaced with $V=s_{xi}s_{xj}$, since the initial x-polarized state is an eigenstate of both operators~\cite{ReyNatPhys2017,ReyPRL2018,SaeedInformScramb}.

In summary, this work verifies that a modified quasi-classical spin vector model of weakly interacting Fermi gases explains perturbed quantum rewinding experiments, using measurements of single-shot spin density profiles with sufficient resolution to enable quantitative study. The modified model reported here elucidates the effects of uncontrolled forward and backward evolution phases, $\varphi_f$ and $\varphi_b$, on the system and measurements, resolving an outstanding conflict with a previous treatment~\cite{SaeedInformScramb}. Our results suggest new correlation analysis methods based on energy-resolved operator products, which yield signals that are independent of the uncontrolled RF detuning without assuming phase distributions~\cite{ThywissenDynamicalPhases}. Applying such methods to measure the time dependence of correlations between transverse components $\bra{\psi_{0x}}\tilde{s}_{\perp i}\cdot\tilde{s}_{\perp j}\ket{\psi_{0x}}$ allows the study of entanglement development in a large system \cite{EntanglePropaNature} and investigations of many-body dynamics and information propagation\cite{PhysRevLett.111.207202}. Such experiments will be a topic of future work.

Primary support for this research is provided by the Air Force Office of Scientific Research (FA9550-16-1-0378). Additional support is provided by the National Science Foundation (PHY-2006234).

$^*$Corresponding authors: jhuang39@ncsu.edu\\ \hspace*{1.65 in} jethoma7@ncsu.edu


%

\widetext
\setcounter{figure}{0}
\setcounter{equation}{0}
\renewcommand{\thefigure}{S\arabic{figure}}
\renewcommand{\theequation}{S\arabic{equation}}

\appendix
\section{Supplemental Material}

This supplemental material presents the experimental and theoretical details of the measurements and modeling of quantum rewinding in a weakly interacting Fermi gas. A new method is introduced for calibration of the magnetic field where the scattering length vanishes. With this precise measurement, systematic experimental defects in the perturbed quantum rewinding experiments are minimized. The critical role of time-dependent detuning is explained by deriving a new collective spin vector model that properly includes it. Finally, additional single-shot data from the perturbed quantum rewinding experiments are presented, illustrating excellent agreement with the quasi-classical model reported in this work.

\subsection{Experimental Methods}
\label{sec:experiment}

For the experiments presented in this work, the sample comprises a mixture of two lowest hyperfine states $^6$Li atoms, denoted $\ket{1}\equiv\ket{\uparrow_z}$ and $\ket{2}\equiv\ket{\downarrow_z}$, which is evaporatively cooled to degeneracy near the $1-2$ broad Feshbach resonance at a bias magnetic field of $832.2$ G. After the sample is prepared, $\ket{1}$ is eliminated by an imaging pulse applied in the weakly interacting regime near $1200$ G to create a $z$-polarized sample. The bias magnetic field is then tuned close to the zero-crossing $527.150$ G, where the scattering length $a$ nearly vanishes. At this field, a resonant radio-frequency(RF) $(\pi/2)_y$ pulse, i.e., a rotation around the $y$-axis in the RF frame, coherently excites the spin state $\ket{2}$, creating a $50$-$50$ superposition of states $\ket{1}$ and $\ket{2}$. In addition to the bias magnetic field, a control magnetic field is applied along the bias field axis by a pair of auxiliary magnet coils, which are wrapped around the primary bias magnet containers, located on the top and bottom of the experimental vacuum chamber. The auxiliary coils enable magnetic field control of the scattering length in the zero crossing region.

With the coherent superposition state created, the trapped cloud is $x$-polarized and evolves at the chosen scattering length $a$ for a selected evolution time $t_{fk}$, after which the spatial profiles of both hyperfine components are measured. Spin wave formation leads to {\it spin segregation}, where the maxima in the $|1\rangle$ and $|2\rangle$ densities are spatially separated \cite{DuSpinSeg1}. This initial evolution is defined as ``forward" evolution. To implement "backward" evolution, a $\pi_y$ pulse is applied after a forward evolution time $\tau_f$, interchanging the two spin states, and the auxiliary magnetic field is swept down over 10 ms to flip the scattering length from $a$ to $-a$. Ideally, from this point, the sign of the Hamiltonian is inverted, which is equivalent to letting the system evolve backward in time. After a backward evolution time $\tau_b=\tau_f$, the system is expected to evolve back to its unsegregated $x$-polarized state.

\subsubsection{Quantum Rewinding Measurements}

The system status is observed for a number of different forward evolution times $t_{fk}$ and backward evolution times $t_{bk}$, by imaging the density of both spins, using two resonant optical pulses, separated in time by $10\,\mu$s. Spin density spatial profiles are extracted by subtracting normalized spatial profiles for the two states and dividing by 2:

\begin{equation}
 S_z(x) \equiv \frac{1}{2}\left(\frac{n_1(x)-n_2(x)}{n_1(0)+n_2(0)}\right)
 = \frac{1}{2}\frac{\Delta n(x)}{n_{tot}(0)}.
\end{equation}
Spin segregation is quantified by measuring the central spin density $S_z(x=0) = \frac{1}{2}(\frac{\Delta n(0)}{n_{tot}(0)})$ at the center of the cloud. The larger the absolute value of the central spin density is, the more segregated the system is: For the unsegregated system, i.e., immediately after the coherent excitation RF $(\pi/2)_y$ pulse, the central spin density is zero in theory.

Perturbed quantum rewinding experiments are done by adding a perturbing $\phi_x$ pulse before the $\pi_y$ pulse. This pulse is generated by connecting a voltage-controlled phase shifter in series with the output of the RF generator, so that a $\phi_y$ pulse can be either unshifted in phase or phase shifted by $90^{\circ}$ to obtain a $\phi_x$ pulse. Immediately after the perturbing $\phi_x$ pulse, the sign of the Hamiltonian is reversed, as described in the previous paragraph, and the system evolves backward for a time period of $\tau_b=\tau_f$. Just before the final $(\pi/2)_y$ pulse, the magnetic field is swept with the auxiliary coils to give the original scattering length $a$, and the final $(\pi/2)_y$ pulse is applied to observe the transverse components of the spin vectors.

\subsubsection{Detuning}

Ideally, during the experimental cycle, for zero detuning, the Bloch frame overlaps with the RF frame, which means that the $\phi_x$, $\pi_y$ and the last $(\pi/2)_y$ rotations in the RF frame are also done about the x-axis or y-axis in the Bloch frame. However, there are uncontrolled time-dependent global detunings $\Delta(t)$, producing relative angles between the RF and Bloch frames during the evolution periods $\tau_b$ and $\tau_f$. Experimentally, all rotations are done in the RF frame, which is defined by the RF generator. At a forward evolution magnetic field $B_f$, which is near resonance, $\Delta\simeq 0$. However, as the magnetic field is swept down to $B_b$ during an experimental cycle, the detuning changes by up to several kHz. This results in a large phase difference and corresponding angle between the two frames, which is imperfectly controlled due to field fluctuations. As described in detail in \S~\ref{sec:model}, the phase shift accumulated during the forward evolution period, $\varphi_f$, controls the effective rotation axis of the perturbation relative to the Bloch vector. In addition, the difference between $\varphi_f$ and the phase accumulated during the backward evolution period, $\varphi_b$, determines the measured spin components.

For Hamiltonian reversal experiments, when only the $z$-component of the spin vectors is measured, there is no sensitivity to the detuning, because the detuning is equivalent to a rotation about the $z$-axis. In contrast, for perturbed quantum rewinding experiments, where the transverse components of the spin vectors are measured, understanding the roles of $\varphi_f$ and $\varphi_b$ is critical for correct data analysis and comparison with predictions.
Although the detuning is not controlled, the experimental results of this work show that complex single-shot data are surprisingly well fitted by the model described below, where the different accumulated phase shifts for the two evolution periods, $\varphi_f$ and $\varphi_b$, are properly included.

\subsection{Collective Spin Vector Evolution Model}
\label{sec:model}

\begin{figure}[htb]
\begin{center}\
\hspace*{-0.0in}\includegraphics[width=3.50in]{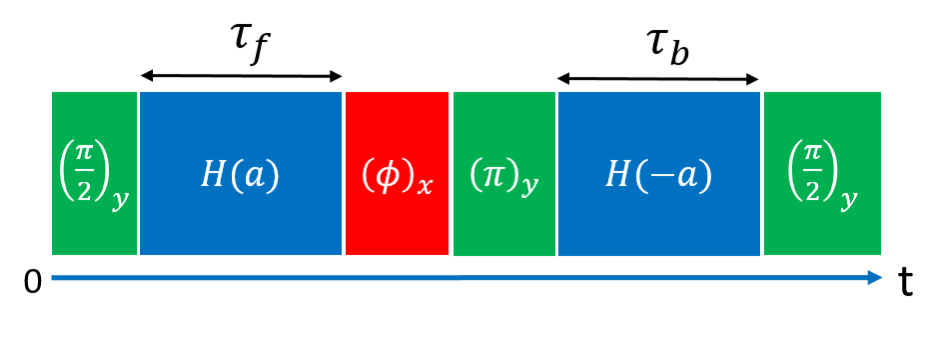}
\end{center}
\caption{Pulse sequence for perturbed quantum rewinding.
\label{fig:OTOCS}}
\end{figure}

To understand how the time-dependent global detuning $\Delta(t)$ affects the perturbed quantum rewinding measurements, we derive the final state of the system after the pulse sequence of Fig.~2 of the main text, which is reproduced here for convenience, Fig.~\ref{fig:OTOCS}.
Prior to the pulse sequence, the optically trapped atoms are initially prepared in a $z$-polarized state,
\begin{equation}
\ket{\psi_{0z}}=\Pi_i\,\ket{\!\!\uparrow_{zi}}.
\end{equation}
The system runs forward for a time $\tau_f$ and backward for $\tau_b$, after which the spatial profiles of the $|\uparrow_z\rangle$ and $|\downarrow_z\rangle$ states are measured. Note that the pulse durations for the $y$- and $x$- axis rotations are $<<\tau_f,\,\tau_b$. In between the forward and backward evolutions, a perturbing pulse is applied, which rotates the system about the $x$-axis by an angle $\phi_x$.

In Fig.~\ref{fig:OTOCS}, the $x$ and $y$ axes are defined in the ``RF frame," where the $x$-axis of the RF frame is defined to rotate about the $z$-axis at the instantaneous frequency of the RF generator. The $x$-axis, therefore, tracks the total phase of the RF field. The detuning $\Delta(t)$ is defined as the difference between the instantaneous hyperfine resonance frequency for an atom at rest and the instantaneous radiofrequency. When the hyperfine frequency is larger than the radiofrequency, spin vectors will appear to rotate counterclockwise as seen looking down the z-axis from above, i.e., through a positive angle, relative to the RF frame. In the experiments, changes in the applied bias magnetic field, as used to reverse the scattering length, and uncontrolled magnetic field fluctuations, tune the hyperfine frequency at a rate of $\simeq 5$ kHz/G. The detuning causes the components of spin vectors in the Bloch frame to rotate relative to the RF-frame by generally different angles, during the forward and backward evolution times, respectively, even for $\tau_b=\tau_f=\tau$ as used in the experiments. We define the forward and backward phase shifts,
\begin{equation}
\varphi_f=\int_{\tau_f}\!\! dt\,\Delta(t)\hspace{0.25in} {\rm and}\hspace{0.25in} \varphi_b=\int_{\tau_b} \!\!dt\,\Delta(t).
\label{eq:1.6}
\end{equation}

To find the final state including the global detuning, we write the Hamiltonian of Eq.~1 of the main text in the general form
\begin{equation}
\frac{H(a)}{\hbar}= \frac{H_0(a)}{\hbar}+\Delta(t)S_{z},
\label{eq:1.1a}
\end{equation}
where $S_z$ is the $z$-component of the dimensionless {\it total} spin vector.
Here, the time-{\it independent} part of the Hamiltonian, for $\Delta=0$, is defined by
\begin{equation}
\frac{H_0(a)}{\hbar}=a\!\sum_{i,j\neq i}\!g_{ij}\,{\vec s}_i\cdot{\vec s}_j+\sum_{i}\Omega'E_i\,s_{zi}
\label{eq:1.1b}
\end{equation}
and
\begin{equation}
[H_0(a),S_z]=0.
\label{eq:commute}
\end{equation}

Referring to Fig.~\ref{fig:OTOCS}, for measurements of spin components in the RF frame, we see that the final state of the cloud for $\tau_f=\tau_b=\tau$ is
\begin{equation}
\ket{\psi_f}=e^{-i\pi/2\, S_y}e^{-\frac{i}{\hbar}\,H_0(-a)\tau-i\varphi_b S_z}e^{-i\pi\, S_y}e^{-i\,\phi_x S_x}e^{-\frac{i}{\hbar}\,H_0(a)\tau-i\varphi_f S_z}e^{-i\pi/2\, S_y}\ket{\psi_{0z}}.
\label{eq:finalstate}
\end{equation}
Eq.~\ref{eq:finalstate} is readily simplified using $$e^{-\frac{i}{\hbar}\,H_0(-a)\tau-i\varphi_b S_z}e^{-i\pi\, S_y}=e^{-i\pi\, S_y}\left[e^{i\pi\, S_y}e^{-\frac{i}{\hbar}\,H_0(-a)\tau-i\varphi_b S_z}e^{-i\pi\, S_y}\right].$$
Using Eq.~\ref{eq:1.1b} and noting that the $(\pi)_y$ rotation inverts $S_z$, we see that $$ e^{i\pi\, S_y}\left[\frac{H_0(-a)}{\hbar}\tau+\varphi_b\,S_z\right]e^{-i\pi\, S_y}=-\frac{H_0(a)}{\hbar}\tau-\varphi_b\,S_z.$$
With Eq.~\ref{eq:commute}, we obtain
\begin{equation}
\ket{\psi_f}=e^{-i 3\pi/2\, S_y}e^{+i\varphi_b S_z}e^{+\frac{i}{\hbar}\,H_0(a)\tau} e^{-i\,\phi_x S_x}e^{-i\varphi_f S_z}e^{-\frac{i}{\hbar}\,H_0(a)\tau}e^{-i\pi/2\, S_y}\ket{\psi_{0z}}.
\label{eq:finalstate2}
\end{equation}

Now, $$e^{-i\,\phi_x S_x}e^{-i\varphi_f S_z}=e^{-i\varphi_f S_z}\left[e^{+i\varphi_f S_z}e^{-i\,\phi_x S_x}e^{-i\varphi_f S_z}\right].$$
It is easy to show that
\begin{equation}
S_x(\varphi_f)\equiv e^{+i\varphi_f S_z} S_x\,e^{-i\varphi_f S_z}=S_x\cos\varphi_f-S_y\sin\varphi_f,
\label{eq:Sxofvarphi}
\end{equation}
which follows from $S_x''(\varphi_f)=-S_x(\varphi_f)$ and the initial conditions $S_x(0)=S_x$ and $S_x'(0)=-S_y$. Then
$$e^{-i\,\phi_x S_x}e^{-i\varphi_f S_z}= e^{-i\varphi_f S_z} e^{-i\phi_x S_x(\varphi_f)}.$$

Again using Eq.~\ref{eq:commute}, we then obtain
\begin{equation}
\ket{\psi_f}=e^{-i 3\pi/2\, S_y}e^{+i(\varphi_b-\varphi_f) S_z}e^{+\frac{i}{\hbar}\,H_0(a)\tau} e^{-i\,\phi_x S_x(\varphi_f)}e^{-\frac{i}{\hbar}\,H_0(a)\tau}e^{-i\pi/2\, S_y}\ket{\psi_{0z}}.
\label{eq:finalstate3}
\end{equation}
Defining the operator
\begin{equation}
W_\phi(\varphi_f,\tau)=e^{\frac{i}{\hbar} H_0(a)\tau}e^{-i\,\phi_x S_x(\varphi_f)}e^{-\frac{i}{\hbar}H_0(a)\tau},
\label{eq:2.9S}
\end{equation}
and the $x$-polarized state just after the first $(\pi/2)_y$ rotation,
\begin{equation}
\ket{\psi_{0x}}=e^{-i\pi/2\, S_y}\ket{\psi_{0z}}
\end{equation}
we obtain the final state in the simple form,
\begin{equation}
\ket{\psi_f}=e^{-i 3\pi/2\, S_y}e^{+i(\varphi_b-\varphi_f)S_z} W_\phi(\varphi_f,\tau)\ket{\psi_{0x}}.
\label{eq:finalstate4}
\end{equation}

As explained in the main text, in a single shot, we can measure the operator $s_{zi}$ for the ensemble of atoms in the i$^{th}$ energy group. Noting that $ e^{+i 3\pi/2\, S_y}s_{zi}e^{-i 3\pi/2\, S_y}=+s_{xi}$, we find
\begin{equation}
\bra{\psi_f}s_{zi}\ket{\psi_f}=\bra{\psi_{0x}}W^\dagger_\phi(\varphi_f,\tau)e^{-i(\varphi_b-\varphi_f)S_z}s_{xi}\,e^{i(\varphi_b-\varphi_f)S_z}W^\dagger_\phi(\varphi_f,\tau)\ket{\psi_{0x}}.
\end{equation}
By using Eq.~\ref{eq:Sxofvarphi} with $\varphi_f\rightarrow \varphi_{fb} \equiv \varphi_f-\varphi_b$, $S_x\rightarrow s_{xi}$ and $S_y\rightarrow s_{yi}$, we see that $$e^{i(\varphi_f-\varphi_b)S_z}s_{xi}\,e^{-i(\varphi_f-\varphi_b)S_z}=s_{xi}\cos\varphi_{fb}-s_{yi}\sin\varphi_{fb}.$$
Then,
\begin{equation}
\bra{\psi_f}s_{zi}\ket{\psi_f}=\bra{\psi_{0x}}W^\dagger_\phi(\varphi_f,\tau)s_{xi}W^\dagger_\phi(\varphi_f,\tau)\ket{\psi_{0x}}\cos\varphi_{fb}-
\bra{\psi_{0x}}W^\dagger_\phi(\varphi_f,\tau)s_{yi}W^\dagger_\phi(\varphi_f,\tau)\ket{\psi_{0x}}\sin\varphi_{fb}
\label{eq:6.4}
\end{equation}

Eq.~\ref{eq:6.4} shows that a single-shot measurement of the spin density profile then yields, via inverse-Abel transformation~\cite{SaeedInformScramb}, the ensemble-averages,
\begin{equation}
\langle\psi_f|s_{zi}|\psi_f\rangle \equiv\bra{\psi_{0x}}\tilde{s}'_{xi}\ket{\psi_{0x}},
\label{eq:6.7}
\end{equation}
where
\begin{equation}
\tilde{s}'_{xi}= \cos\varphi_{fb}\,\tilde{s}_{xi}-\sin\varphi_{fb}\,\tilde{s}_{yi}.
\label{eq:3.9}
\end{equation}
Here,
\begin{equation}
\tilde{s}_{xi}\equiv W^\dagger_\phi(\varphi_f,\tau)s_{xi}W_\phi(\varphi_f,\tau)
\end{equation}
and similarly for $\tilde{s}_{yi}$, reproducing the results given in the main text.

We see that for each measurement, the difference between the forward and backward phase shifts, $\varphi_f-\varphi_b\equiv\varphi_{fb}$, determines the relative contribution of the $\tilde{s}_{xi}$ and $\tilde{s}_{yi}$ spin components in the Bloch frame to the measured projection in the RF frame, $\tilde{s}'_{xi}$. In addition, Eq.~\ref{eq:2.9S} shows that the forward phase shift $\varphi_f$ determines the effective rotation axis for the $\phi_x$ pulse.

To compare the prediction of Eq.~\ref{eq:6.7} to single-shot measurements in a system containing a large number of spins, we employ a quasi-classical model. In this case, we treat the Heisenberg equations for the spin-vectors $\vec{s}_i$ as evolution equations for classical vectors, which neglects quantum correlations. These equations are readily evaluated for any chosen $\varphi_f$ and $\varphi_b$ by numerical integration, enabling fits to single-shot data with $\varphi_f$ and $\varphi_b$ as fit parameters.

\subsection{Quantifying Hamiltonian Reversal}
\label{sec:quantHre}

This section introduces the method of quantifying the quality of rewinding by Hamiltonian reversal.
Two sets of data are required to examine the result of the Hamiltonian reversal in the experiments. One set of data represents the state of the system at different forward evolution times, and the other set represents the state at corresponding backward evolution times. The first set is taken at a magnetic field $B_f$ for a number of times $0\leq t_{fk}\leq\tau_f$, where $t=0$ is the time of the initial coherent excitation pulse and $\tau_f$ is the maximum forward evolution time. The second set is taken at a backward evolution magnetic field $B_b$ at corresponding times $t_{bk}=\tau_f+\tau_{bk}$, as discussed below.


The $z$-component of the spin density for the forward evolving system, $S^f_{z}(x,t_k)$, is measured at different evolution times $t_{fk}\equiv t_k$ by imaging the density of both spins at a time $t_{fk}$ relative to the time $t\equiv 0$ of the initial
$(\pi/2)_y$ RF pulse. To enable a determination of the reversal quality, the same quantity, $S^b_{z}(x,t_k)$, is measured for the Hamiltonian reversed system at $t_{bk}\equiv \tau_f+\tau_{bk}$, where $t_{bk}$ is the total time relative to the coherent excitation pulse. Here, $\tau_{bk}$ is the amount of time that the system evolves backward with the reversed Hamiltonian. To match the spin density spatial profiles for the forward evolution with the corresponding backward evolution ones, the evolution times need to be matched, i.e., $\tau_{bk}=\tau_f-t_{fk}=\tau_f-t_{k}$ and $t_{bk}= 2\tau_f-t_k$. $\chi^2_k$ is defined as the normalized mean square difference between forward and backward evolution spin density profiles for $t_k$,

\begin{equation}
\chi^2_k \equiv \sum_{x} \left(\frac{S^b_{z}(x,t_k)-S^f_{z}(x,t_k)}{S^f_{z}(x,t_k)}\right)^2.
\label{eq:chiSqr}
\end{equation}
To quantify the reversal, we employ $\chi^2\equiv\langle\chi^2_k\rangle$, the average $\chi_k^2$ for all of the $t_k$ in the data set. Small $\chi^2$ means that the forward and backward $S_z(x)$ profiles overlap very well, which corresponds to a good reversal.

\subsubsection{Zero-crossing Measurement}
\label{sec:zeroCrossing}
\begin{figure}[htb]
\begin{center}\
\hspace*{-0.1in}
\includegraphics[width=7in]{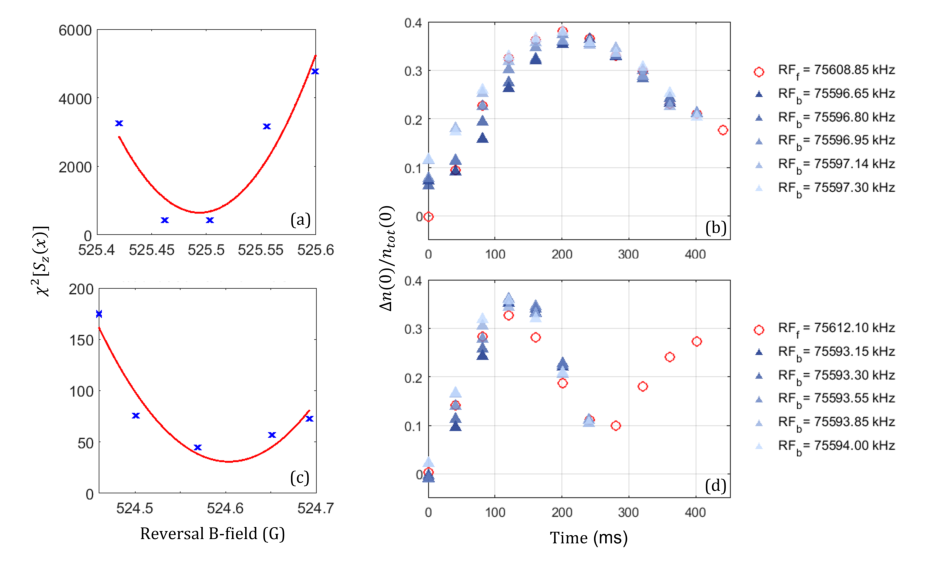}
\end{center}
\caption{Finding the optimum magnetic fields for reversing the spin evolution at two initial magnetic fields. (a)(b) are done at the forward evolution field $B_f = 528.803$ G ($a\approx5\,a_0$), where the resonant RF frequency is $RF_f = 75608.85$ kHz, with the sign of Hamiltonian inverted after forward evolution time $\tau_f = 400$ ms; (c)(d) are done at $B_f = 529.713$ G ($a\approx8\,a_0$), where the resonant RF frequency is $RF_f = 75612.10$ kHz, with the sign of Hamiltonian inverted after $\tau_f = 240$ ms. (a)(c) quantify the reversibility with different reversal magnetic fields $B_b$. Blue crosses are calculated $\chi^2$ results for forward and backward evolution spin density spatial profiles $S_z(x)$. Red curves are parabolas fitted to $\chi^2$ as a function of reversal magnetic field $B_b$. The parabola fits suggest that the optimum reversal field for $B_f = 528.803$ G is $B_{b,opt} = 525.488$ G and the optimum reversal field for $B_f = 529.713$ G is $ B_{b,opt}=524.596$ G. (b)(d) show the central spin density $\Delta n(0)/n_{tot}(0)$ at different evolution times for both forward (red circle) and backward (blue triangles) evolution data. Darker blue corresponds to higher resonant frequency $RF_b$ at $B_b$, which means a higher magnetic field.
\label{fig:zeroCrossingHReversal}}
\end{figure}
A critical constant for the implementation of any experiments involving quantum rewinding is the zero-crossing magnetic field $B_0$ where the scattering length vanishes. Careful calibration of this field ensures that the Hamiltonian reversal in the experimental sequence is done properly. In this work, $B_0$ is precisely measured by quantifying Hamiltonian reversal for two different forward evolution magnetic fields $B_f = 528.803$ G and $529.713$ G, where the scattering lengths are $\approx5\,a_0$ and $\approx8\,a_0$ respectively. Data is taken for five slightly different backward evolution magnetic fields $B_b$ near the zero crossing for each $B_f$.

The magnitudes of all of the magnetic fields are measured precisely using RF spectroscopy, by applying a $\pi$ pulse (15 ms) with a known RF frequency. The resonance frequencies of the RF pulse for the atomic transition are in one-to-one correspondence with the magnetic fields. With this property, the magnetic field can be calculated with mG precision from the resonance RF frequency for a $\pi$ pulse that fully transfers atoms from $|2\rangle$ to $|1\rangle$.

By fitting a parabola to $\chi^2$ as defined above for five different $B_b$, the optimum reversal magnetic $B_{b, opt}$ is obtained for the corresponding $B_f$. The zero-crossing magnetic field is located at the midpoint between $B_f$ and $B_{b,opt}$. Fig.~\ref{fig:zeroCrossingHReversal} displays the results of this measurement. Because the measurement for this experiment is insensitive to the detuning as described in the main paper, averaging is allowed. Each data point is the result after averaging 5 shots. The top two figures (a) and (b) are the result of data series taken for $a=\pm 5.2~a_0$ with Hamiltonian reversal done at $\tau_f=400$ ms, and the bottom two (c) and (d) are the results for $a=\pm 8.0~a_0$ with Hamiltonian reversal done at $\tau_f=240$ ms. The parabolic fit for $\chi^2$ with $a=\pm 5.2~a_0$ series suggests that $B_{b,opt}=525.488$ G for $B_f=528.803$ G, and for $\pm8.0~a_0$ suggests that $B_{b,opt}=524.596$ G for $B_f=529.713$ G. The two series of experiments yield the result $B_0=527.150(5)$ G, which is $30$ mG lower than the previous result~\cite{SaeedPRASpinECorrel}. Note that the scattering lengths ''$5.2\,a_0$" and ''$8.0\,a_0$" are calculated based on the zero-crossing magnetic field measured in the above experiment.

\subsubsection{Reversibility in Different Regimes}
\label{sec:rvsblty}
This technique of quantifying the reversal quality can also be used to compare the reversibility for systems in different regimes of scattering length and forward evolution time. This is achieved by weighting the disagreement between forward and backward in spatial profiles differently for different $t_k$: By construction, the disagreement is weighted more heavily for small $t_k$, where the system has evolved backward long enough for discrepancies between backward and forward evolution to appear. For small $t_k$, $S^f_{z}(x, t_k)$ is usually small, i.e., close to a horizontal line around $x$-axis, because the system has not segregated very much for short forward evolution times. Hence, the denominator in Eqn.~\ref{eq:chiSqr} exaggerates the magnitude of $\chi^2_k$ for small $t_{fk}$.

For the perturbed quantum rewinding experiments, the evolution times are always chosen to be $\tau_b=\tau_f$, which means that the reversal quality at $t_k=0$ ms is extremely critical. Hence, $\chi^2_k$ at $t_k=0$ ms needs to be weighted highly in the test of reversibility, especially for the purpose of choosing regimes to do perturbed quantum rewinding experiments.

 Note that the $\chi^2$ in Fig.~\ref{fig:zeroCrossingHReversal}(a) is one order of magnitude larger than it is in (c), which means that the system reverses more perfectly with $a=8.0~a_0$ and $\tau_f = 240$ ms than with $a=5.2~a_0$ and $\tau_f=400$ ms. This matches the comparisons of the central spin density evolutions shown in Fig.~\ref{fig:zeroCrossingHReversal} (b)(d): For the $a=5.2~a_0$ and $\tau_f=400$ ms data series, there is a clear sign of segregation ($\Delta n(0)/n_{tot}(0)>0$) for Hamiltonian reversed data at $t_k=0$ ms even if the optimum reversal magnetic field is adopted. In contrast, for $a=8.0~a_0$ and $\tau_f=240$ ms, an almost perfect reversal is observed at $t_k=0$ ms.

With this method of quantifying quantum rewinding, a systematic study of the reversibility of a system in different regimes can be done by fixing $\tau_f$ and varying the scattering length and by fixing the scattering length and varying $\tau_f$. This study is ongoing and requires a large amount of data, and so will not be pursued further here.

\subsection{Testing the Quasi-Classical Spin Model}
To test the quasi-classical spin model, as reported in this work, three series of perturbed quantum rewinding experiments are performed at: $5.2\,a_0$ with $\tau=200$ ms, $8.0\,a_0$ with $\tau=200$ ms, and $5.2\,a_0$ with $\tau=400$ ms. The $S_z(x)$ profiles are measured as described in \S~\ref{sec:experiment}. All spatial profiles shown in this work are folded over the center of the cloud, $x=0$, followed by equal-width binning into 50 bins. To extract energy-space information, $S_z(E)$, Abel inversion is applied to the spatial profiles with 16 expansion terms~\cite{NewAbelInversion}. As energy-space profiles are less sensitive to experimental defects shown in Fig. \ref{fig:wrongImbalance}, the model is fitted to the $S_z(E)$ extracted from single-shot data with $\varphi_f$ and $\varphi_b$ as free parameters. Data from the three experimental series are in good agreement with the model.

\subsubsection{Primary Data for $a=5.2~a_0$ and $\tau=200$ ms }
\label{sec:SuppMainDataSet}
\begin{figure}[htb]
\begin{center}\

\includegraphics[width=6.5in]{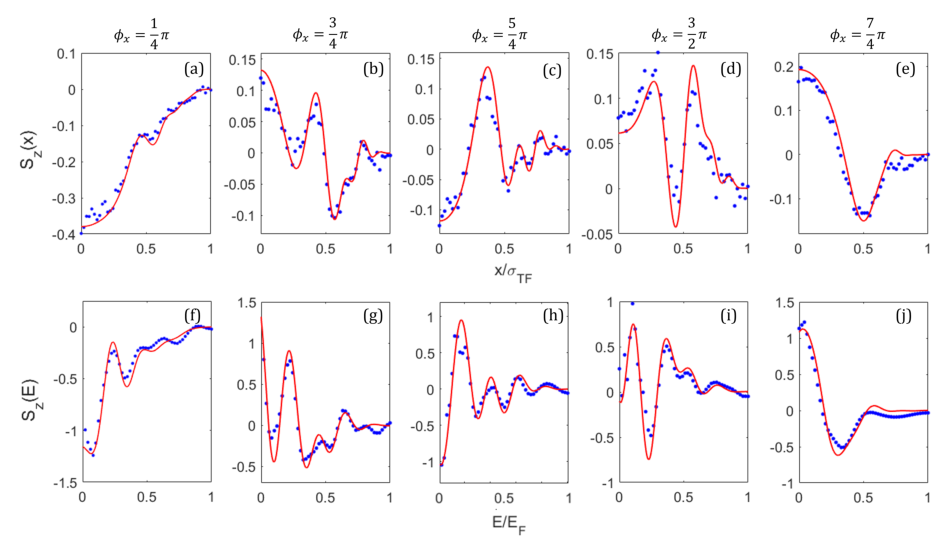}
\end{center}
\caption{Additional single-shot spin density profiles for perturbed quantum reversal experiment at $5.2~a_0$, $\tau=200$ ms, with different $\phi_x$. Blue dots in (a)(b)(c)(d) are measured single shot $S_z(x)$ and those in (e)(f)(g)(h) are Abel inversion of (a)(b)(c)(d), which give $S_z(E)$. Red curves are model fitting data with $\varphi_f$ and $\varphi_b$ as two free parameters. Perturbation rotation angle about the x-axis $\phi_x$ is: $\pi/4$ in (a)(f), $3\pi/4$ in (b)(g), $5\pi/4$ in (c)(h), $3\pi/2$ in (d)(i), $7\pi/4$ in (e)(j). $\phi_x=\pi/2$ and $\pi$ examples are shown in the main paper, Fig.~4.
\label{fig:allProfiles}}
\end{figure}

Perturbed quantum rewinding measurements are employed to precisely test the quasi-classical spin model presented in \S~\ref{sec:model}. Data is mainly taken at $5.2~a_0$ with $\tau=200$ ms, as this set of parameters is expected to provide an almost perfect Hamiltonian reversal. Nine different $\phi_x$ values are used as a perturbation to the reversal, ranging from $0$ to $2\pi$ in steps of $\pi/4$. Fig.~\ref{fig:allProfiles} shows examples of single shot data taken for different $\phi_x$. Note that in these figures $\varphi_f$ and $\varphi_b$ are not necessarily the same across the whole data set because of fluctuations in RF detuning as discussed in \S~\ref{sec:model}. The systematic measurements and fits done for the perturbed quantum rewinding experiments show that the complicated structure observed in the spatial profiles is very sensitive to the initial conditions (cloud size $\sigma_{TF}$ and atom number $N$), as well as the RF detunings for two evolution periods $\tau_f$ and $\tau_b$. Even small variations in these parameters result in a slightly shifted/skewed spatial profiles. Hence, for the purpose of quantitative tests of the quasi-classical spin model using perturbed quantum rewinding experiments, as presented in this work, {\it single-shot} analysis is essential for processing data with imperfectly controlled experimental parameters, which tends to wash out the fine structure in the spatial profiles. The measured single-shot profiles presented in this work have adequate spatial resolution to capture small details in the profiles. Having minimized experimental defects by careful calibrations, the measured single-shot data provide stringent tests of predictions based on the model of \S~\ref{sec:model}.

\subsubsection{Additional Data Sets}
In this section, we present additional data for increased scattering length and evolution time.
The validity of the modified quasi-classical spin model of \S~\ref{sec:model} is demonstrated for the primary data set with $a=5.2~a_0$ and $\tau=200$ ms in \S~\ref{sec:SuppMainDataSet}. To test the model further, a series of additional perturbed quantum rewinding experiments are done with $\tau=200$ ms and $8.0~a_0$ and with $\tau=400$ ms and $a=5.2~a_0$ for three $\phi_x$ values: $\pi/2$, $\pi$ and $3\pi/2$. Less quantitative agreement between the model and data is observed in many of the single shots from these additional experiments, especially in the spatial profiles.

\begin{figure}[htb]
\begin{center}\
\hspace*{-0.1in}
\includegraphics[width=7in]{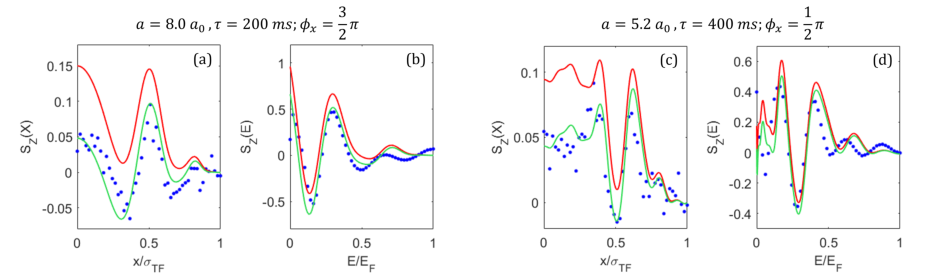}
\end{center}
\caption{Adjusting the atom number ratio for the two spin states in the model results in a much better agreement with data. In (a)(c), blue dots are measured single-shot spin density $S_z(x)$ with $8.0~a_0$, $\tau=200$ ms, $\phi_x = 3\pi/2$ and $5.2~a_0$, $\tau=400$ ms, $\phi_x = \pi/2$ respectively. Red curves are generated by the model assuming a perfectly stable $B$-field, so that all RF pulses are on resonance. Green curves are obtained by adjusting the spin imbalance of the two spin states as described in the text: $n_1(x) \rightarrow n_1(x)-\delta n_{tot}(x)$ and $n_2 \rightarrow n_2(x)+\delta n_{tot}(x)$, with $n_1(x)$ and $n_2(x)$ from the red curve model. $\delta=10\%$ for (a) and $\delta=5\%$ for (c). (b)(d) are Abel inversion of (a)(c) with 16 cosine terms, with color coding the same as (a)(c).
\label{fig:wrongImbalance}}
\end{figure}

The disagreement appears from differences between the spin imbalances in the model and in the data. In Fig.~\ref{fig:wrongImbalance}, the basic model (red curves) assumes that all of the applied RF pulses are on resonance, while drifts in the detuning alter the pulse area. It is clear that the basic model still captures the shape and oscillations of the data (blue dots), but with an offset. In the experiments with longer experimental cycles or larger magnetic field sweeps, there is a larger probability that one or more of the RF pulses are slightly off resonance with the hyperfine frequency, since the RF frequency is fixed for all pulses, but the magnetic field can fluctuate because of the limitation on the stability of the auxiliary coils. Imperfect RF pulses result in a measurement with the wrong spin imbalance for the given $\phi_x$, compared to ideal measurements. To include the effect of imbalance in the model, the atom numbers for the two spin states are adjusted from $N_1$ and $N_2$ to $N_1-\delta N_{tot}$ and $N_2+\delta N_{tot}$, with $N_{tot} = N_1+N_2$ being total atom number and $\delta$ a reasonably small ($\leq 10$\%) fraction. In our system, the total density $n_{tot}(x) = n_1(x)+n_2(x)$ is invariant during the experimental cycle. Hence, to include the adjustment of spin imbalance in the modeled spatial profile, the outcomes $n_1(x)$ and $n_2(x)$ are first determined from the model and then scaled to $n_1(x)-\delta n_{tot}(x)$ and $n_2(x)+\delta n_{tot}(x)$. In this way, the total atom number and density profile remain the same in the model output before and after the adjustment. With this adjustment to the outcome of the regular model, the improved fits shown as the green curves in Fig.~\ref{fig:wrongImbalance} are obtained.

Another reason for the disagreement between the model and the data set with $\tau=400$ ms at $a=5.2~a_0$ is the imperfect Hamiltonian reversal shown in Section~\ref{sec:zeroCrossing}. The unperturbed quantum rewinding experiment done with this set of experimental parameters suggests that the system is not precisely reversed in this regime. Therefore, it is reasonable that the perturbed rewinding data can not be predicted by the model as quantitatively as for the data obtained in the regime where the reversibility of the system is clearly better.


\end{document}